# Efficient and tunable narrowband second-harmonic generation by a large-area etchless lithium niobate metasurface

Yaping Hou[1], Yigong Luan[2], Yu Fan[3], Alfonso Nardi[2], Attilio Zilli[2], Bobo Du[1], Jinyou Shao[3], Marco Finazzi[2], Chunhui Wang[3]*, Lei Zhang[1]*, Michele Celebrano[2]*

[1]*Key Laboratory of Physical Electronics and Devices of Ministry of Education & Shaanxi Key Laboratory of Information Photonic Technique, School of Electronic Science and Engineering, Xi'an Jiaotong University, Xi'an, 710049, People's Republic of China*
[2]*Department of Physics, Politecnico di Milano, 20133, Milano, Italy*
[3]*Micro- and Nano-Technology Research Center, State Key Laboratory for Manufacturing Systems Engineering, Xi'an Jiaotong University, Xi'an, 710049, People's Republic of China.*

*Correspondence author: *Michele Celebrano, Lei Zhang, Chunhui Wang*
*Yaping Hou, Yigong Luan, and Yu Fan* contribute equally to this work.

Optical resonances in nanostructures enable strong enhancement of nonlinear processes at the nanoscale, such as second-harmonic generation (SHG), with high-Q modes providing intensified light–matter interactions and sharp spectral selectivity for applications in filtering, sensing, and nonlinear spectroscopy. Thanks to the recent advances in thin-film lithium niobate (TFLN) technology, these key features can be now translated to lithium niobate for realizing novel nanoscale nonlinear optical platforms. Here, we demonstrate a large-area metasurface, realized by scalable nanoimprint lithography, comprising a slanted titanium dioxide ($TiO_2$) nanograting on etchless TFLN for efficient narrowband SHG. This is enabled by the optimal coupling of quasi-bound state in the continuum (q-BIC) modes with a narrowband pulsed laser pump. The demonstrated normalized SHG efficiency is 0.15% $cm^2$/GW, which is among the largest reported for LN metasurfaces. The low pump peak intensity (3.64 kW/cm²) employed, which enables SHG even by continuous-wave pumping, allows envisioning integrated and portable photonic applications. SHG wavelength tuning from 870 to 920 nm with stable output power as well as polarization control is also achieved by off-normal pump illumination. This versatile platform opens new opportunities for sensing, THz generation and detection, and ultrafast electro-optic modulation of nonlinear optical signals.

**Keywords:** second-harmonic generation, etchless metasurface, thin-film lithium niobate,

dielectric nanophotonics, quasi-bound state in the continuum

**Introduction**

The rapid advancement in integrated photonics has driven a strong demand for miniaturized and energy-efficient nonlinear platforms to modulate, manipulate, and shape the wavefront of light. The generation of optical harmonics is an archetypal nonlinear optical process that has been translated to the nanoscale[1,2] to enable applications ranging from ultrafast signal processing[3,4] to biosensing,[5] and nonlinear imaging.[6] However, the conversion efficiency of nonlinear optical processes drops significantly in ultra-compact volumes. Such limitation can be mitigated by exploiting optical resonances in nanostructured materials, such as nanoparticles and metasurfaces, that offer strong field enhancement and spatial confinement, hence boosting the frequency conversion process.[7,8] Specifically, this can be achieved by leveraging three phenomena: the electromagnetic field enhancement at both the fundamental and harmonic frequencies, the spatial overlap of the interacting modes, and the intrinsic second-order nonlinear susceptibility of the medium mediating the photon upconversion.[9] Engineering nanostructures that simultaneously optimize these mechanisms is therefore crucial for realizing efficient, on-chip nonlinear optical functionalities.

Metallic nanostructures have been widely explored thanks to their ability to concentrate light into nanometric hotspots through plasmonic resonances,[10,11] but their applicability is fundamentally constrained by strong Ohmic losses and low damage thresholds. Therefore, interest has recently shifted to dielectric materials, particularly to those with a crystal lattice with broken inversion symmetry that allow for second-harmonic generation (SHG) in the bulk.[12] Among these, lithium niobate ($LiNbO_3$, LN) emerged as a prime materials thanks to its large second-order nonlinearity, wide transparency window, and high damage threshold. Recent advances in the fabrication of LN thin films underpinned the demonstration of nonlinear optical functionalities in ultracompact platforms, such as photonic crystal, microdisk resonators, and waveguides,[13,14] including metasurfaces enabling SHG.[15,16]

In this context, dielectric metasurfaces and nanogratings featuring nonlocal photonic modes, such as quasi-bound states in the continuum (q-BICs), were found capable to provide a strong field enhancement,[17] which is beneficial to nonlinear optical applications[18] including

SHG.[19–21] However, the fabrication of high-quality large-surface area LN nanostructured surfaces, capable of sustaining delocalized photonic modes, is notoriously challenging, because of the high chemical stability of LN, which makes it difficult to realize smooth and vertical sidewalls with conventional etching techniques. Moreover, the direct etching of LN is well known to be affected by material redeposition and induce defects in the crystalline structure, hence hindering the nonlinear conversion efficiency even by orders of magnitude.[22] New approaches to nanofabrication have circumvented this issue through the development of etchless LN platforms, where low-refractive-index periodic structures are patterned atop a thin-film lithium niobate (TFLN) layer, thereby avoiding the need to directly etch LN itself. These platforms already demonstrated the ability to support nonlocal photonic modes, such as q-BICs, exhibiting extremely high quality factors (Q-factors).[23–25]

In this work, we circumvent the major challenges in LN nanofabrication by leveraging nanoimprint lithography to realize an etchless LN platform for efficient, narrowband, and tuneable SHG. The device consists of a large-area (1 mm × 1 mm) titanium dioxide ($TiO_2$) nanograting fabricated on top of an $x$-cut LN film by nanoimprint lithography. Under near-normal incidence pumping, the nanograting provides strong field localization within the $LiNbO_3$ film via nonlocal q-BICs obtained by breaking the mirror symmetry via slanting the $TiO_2$ grating. The use of a picosecond pulsed laser source in place of the more usual femtosecond oscillators allows to achieve optimal matching with the q-BIC resonance linewidth, hence maximizing the power in-coupling into the optical mode. This design combines high nonlinear susceptibility, strong resonant field enhancement, and scalable fabrication, enabling narrowband SHG without directly etching the LN crystal.

**Results and discussion**

Figure 1a shows a photograph of the fabricated sample having a size of 1 mm × 1 mm. The device consists of an array of slanted $TiO_2$ nanowires – forming a one-dimensional nanograting – patterned by nanoimprint lithography on an $x$-cut TFLN supported by a silicon dioxide ($SiO_2$) substrate. The nanowires run parallel to the optical ($z$) axis, as shown in Fig.1b. They are $W$ = 530 nm wide with periodicity $P$ = 910 nm and a slanting angle $\alpha$ = 5°, while the $TiO_2$ and TFLN thicknesses are respectively 510 nm and 610 nm, based on the scanning electron microscopy image in Fig. 1c.

To engineer the optical modes of the device in the region of interest, we first calculated the photonic band structure of the system with vertical sidewalls ($\alpha = 0°$), from which four resonant modes can be identified: $TE_{10}$ and $TE_{20}$ for TE-polarized plane-wave illumination (with the optical electric field polarized along the $z$-axis) and $TM_{20}$ and $TM_{10}$ for TM-polarized plane-wave illumination (with the optical electric field polarized along the $x$-axis), as shown in Fig. 1d. The optical modes were investigated using finite-element method as implemented in the commercial solver COMSOL Multiphysics, and the detailed simulation procedure is described in the Methods section. Because of mirror symmetry of the unit cell, at the avoided crossing point the two guided modes split into a leaky mode (LM) and a dark one, commonly known as a symmetry-protected bound state in the continuum (BIC).[30] We have then evaluated the $Q$-factor of each optical mode as a function of the grating slanting angle $\alpha$ (see Fig. 1e). The $Q$-factors of the modes $TE_{10}$ and $TM_{10}$ exhibit a steep dependence on the slanting angle, which diverges at $\alpha = 0°$. In contrast, the modes $TE_{20}$ and $TM_{20}$ feature a moderately high $Q$-factor largely insensitive to $\alpha$. While the first behavior is a signature of BIC resonances, the second one is typical of LM resonances. The fabricated structure has been designed with a slanting angle $\alpha = 5°$ to perturb the mirror symmetry, hence converting the BIC into a q-BIC with a theoretical $Q$-factor of about $10^5$. This value has been selected to provide large field enhancement in the nonlinear LN film while maintaining an efficient coupling to a collimated free-space pump beam. To maximize the energy delivered to the sample, a narrowband picosecond laser with an equivalent $Q$-factor larger than $10^3$ is employed. The spatial field distribution maps for $\alpha = 5°$ are shown in Fig. 1f, indicating that all four modes are strongly localized within the TFLN layer, albeit with different degrees of confinement and spatial distribution in the layer.

Figure 2 reports the simulated and experimental transmission spectra of the nanograting as a function of the angle $\theta$ between the sample normal and the pump beam for both TE (panels a and b) and TM (panels c and d) excitation. The experimental transmission spectra, acquired with a supercontinuum laser source by tilting the sample around the $y$ axis by means of a motorized rotation stage from −4° to +4° in steps of 0.2°, are in excellent agreement with the simulations and closely follow the dispersion bands displayed in Fig. 1d. While the $Q$-factor of the modes TE$_{10}$ and TM$_{10}$ show a marked dependence on the incidence angle (sample rotation angle $\theta$) – a clear signature of q-BICs – LMs TE$_{20}$ and TM$_{20}$ exhibit broader resonances with linewidths largely insensitive to the tilting angle $\theta$, consistently with the simulations shown in Fig. 1e. Notably, the q-BICs (TE$_{10}$ and TM$_{10}$) and LMs (TE$_{20}$ and TM$_{20}$) exhibit dispersion with opposite slopes as a function of sample rotation angle $\theta$ (see Figs. 2b and 2d).

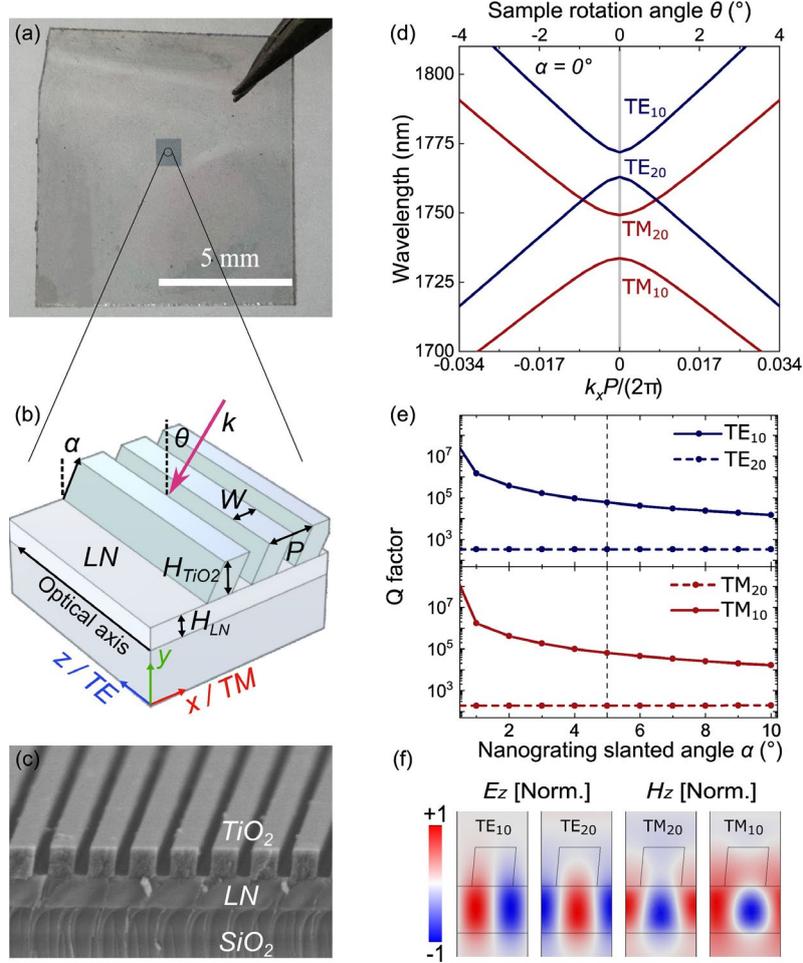

**Figure 1.** Structural characterization and optical response of the slanted TiO$_2$ grating on TFLN. (a) Photograph of the sample. The patterned area is 1 mm × 1 mm. (b) Schematic image of the TiO$_2$ nanograting on a LiNbO$_3$ film. The optical axis of LiNbO$_3$ is along the z-axis. The pump beam impinges on the device from the nanograting side, while the second-harmonic signal is collected from the side of the substrate. (c) Cross-sectional scanning electron micrograph of the fabricated sample. The TiO$_2$ nanowires (with period $P$ = 910 nm, width $W$ = 530 nm, height $H_{TiO2}$ = 510 nm, and tilt angle $\alpha$ = 5°) are arranged on a LiNbO$_3$ layer of thickness $H_{LN}$ = 610 nm. (d) Mode dispersion in the reciprocal space of the TiO$_2$ nanograting with $\alpha$ = 0°. (e) Grating tilt-angle dependence of the extracted quality factors for TE$_{10}$, TE$_{20}$, TM$_{20}$ and TM$_{10}$ modes. (f) Spatial distribution of the dominant component of the electric (magnetic) field distributions for the TE (TM) modes in a single nanograting unit cell (with $\alpha$ = 5°) for normal incidence excitation.

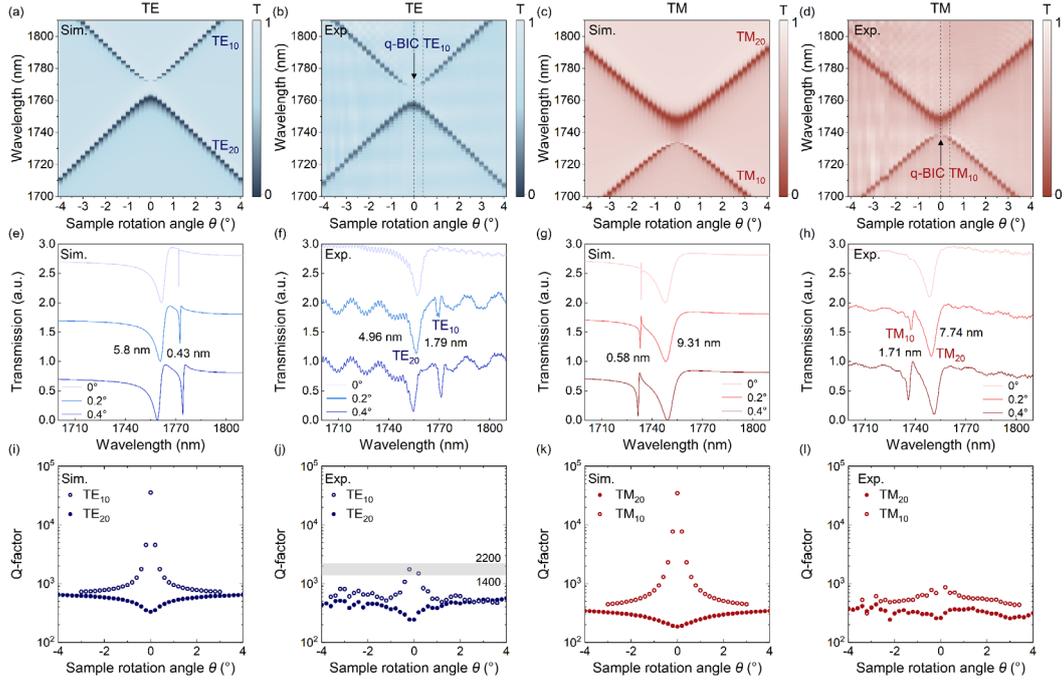

**Figure 2.** Angular dispersion of the transmission spectra and Q-factor of the slanted TiO$_2$ grating on TFLN under TE- and TM-polarized excitations. (a,b) Simulated and experimental transmission spectra with respect to the sample rotation angle under TE-polarized excitation. (c,d) Simulated and experimental transmission spectra with respect to the sample rotation angle under TM-polarized excitation. The dashed vertical lines in the panels (b,d) mark the angles $\theta$ of 0° and 0.4°, respectively. (e,f) Simulated and experimental transmission spectra under TE-polarized excitation at sample rotation angles $\theta$ of 0°, 0.2°, and 0.4°. (g,h) Simulated and experimental transmission spectra under TM-polarized excitation at sample rotation angles $\theta$ of 0°, 0.2°, and 0.4°. (i,j) Simulated and experimental Q-factor variation as a function of the sample rotation angle $\theta$ under TE-polarized excitation. The blue hollow circles represent the q-BIC TE$_{10}$, while the blue solid circles correspond to the TE$_{20}$ mode. The grey shaded area in (j) indicates the Q-factor range of the ps laser, which spans from 1400 to 2200. (k,l) Simulated and experimental Q-factor variation as a function of the sample rotation angle $\theta$ under TM-polarized excitation. The red solid circles represent the mode TM$_{20}$, while the red hollow circles correspond to the q-BIC TM$_{10}$.

A key feature of this device is that the slanted geometry, in principle, enables coupling to the q-BICs even at normal light incidence. We have thus examined the linewidth evolution by comparing three representative transmission spectra near $k_x = 0$ for both TE (Figs. 2e and 2f) and TM (Figs. 2g and 2h) excitation. At normal incidence ($\theta = 0°$), the q-BICs are clearly resolved in simulations (see Figs. 2e and 2g) but are barely detectable experimentally (see Figs. 2f and 2h), being narrower than the spectrometer resolution (~ 0.5 nm). At tilt angles of $\theta = 0.2°$ and 0.4°, the $Q$-factors of both TE and TM q-BIC modes are reduced to less than 10$^3$. Conversely, LM TE$_{20}$ exhibits a $Q$-factor spanning 250 to 450 (350 to 650) in the experiment (simulation), while LM TM$_{20}$ maintains a nearly constant experimental linewidth (see Fig. 2l) for all tilt angles. The measured linewidths are systematically broader than those predicted by simulations (see Figs 2i-2l). Such broadening can be ascribed to (i) the divergence of the supercontinuum source,[26] which introduces in-plane wave vectors and thus smears the angular resolution; (ii) fabrication imperfections, which may introduce defects that further lower the asymmetry of the guided modes.[27]

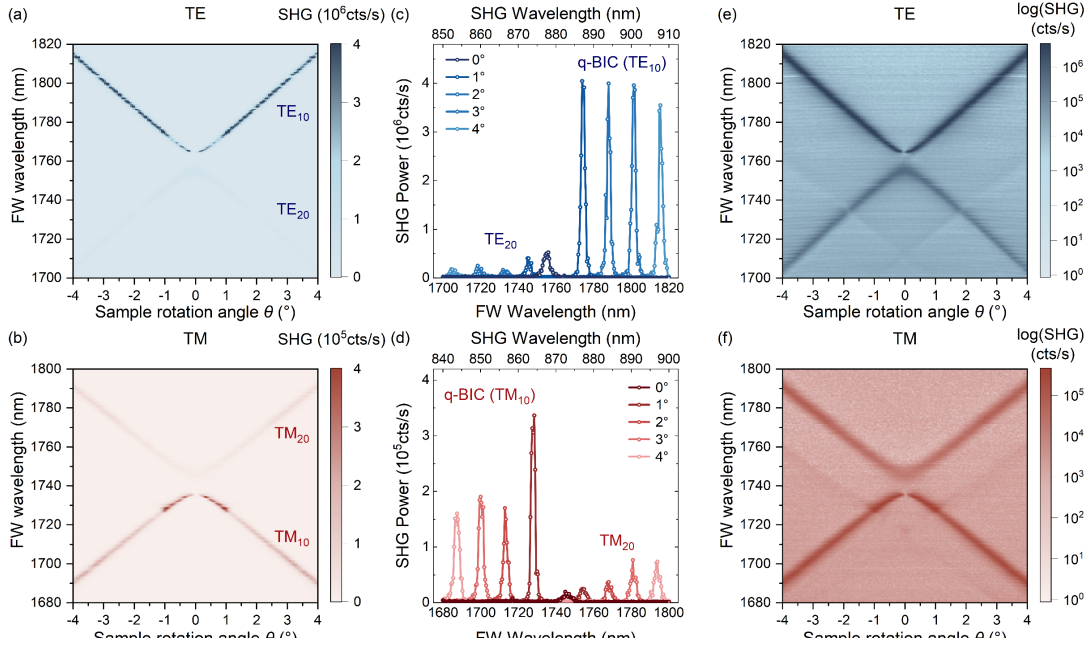

**Figure 3.** SHG emission from the slanted TiO$_2$ grating on TFLN. (a,b) SHG emission spectra as a function of the sample tilt angle under TE- and TM-polarized excitation, respectively. (c,d) Experimental SHG spectra for various tilt angles from 0° to 4° (vertical line cuts of panels a and b), under TE- and TM-polarized excitation, respectively. (e,f) Same as (a,b), but in a logarithmic color scale. Additional features appearing are ascribed to the resonant guided modes (c) TE$_{31}$, TE$_{41}$ and (f) TM$_{31}$, TM$_{41}$ at the SHG wavelength as detailed in Figure S5 of the Supplementary Information (SI).

Building on the identification of q-BIC and LM from the linear transmission study, we also investigated the angle-resolved SHG response from the designed nanograting. As anticipated, to properly couple to the ultranarrow q-BIC resonances and fully exploit the associated field enhancement, we excited the sample with collimated picosecond pulses produced by a narrowband ($\Delta k < 4$ cm$^{-1}$ → $\Delta \lambda < 1$ nm) tunable parametric oscillator. The laser Q-factor, defined as $Q_{laser} = \Delta\lambda/\lambda$, ranges between 1400 to 2200 and was schematized as a grey band in Fig. 2j. It grants, in principle, optimal matching with the q-BICs already at very small angles ($\theta < 1°$, see Fig. 2j). The sample is illuminated from the nanograting side, while the SHG signal is collected from the substrate side. The average power incident on the sample was kept constant with an automated feedback loop for each wavelength and sample tilt angle, with

$P$ = 10 mW for both TE and TM excitations. We carried out a complete characterization of the SHG under both TE- and TM-polarized excitation by scanning the pump wavelength in steps of 0.5 nm and the sample tilt angle in steps of 0.05 degree around the $z$-axis, see Figure 3.

Under TE excitation, the pump beam propagates along the $y$-axis with polarization aligned to the LN optical axis ($z$-axis), hence addressing the dominant nonlinear coefficient $d_{33}$ = 29.1 pm/V (at $\lambda \approx$ 1750 nm).[28] The resulting SHG power map is shown in Fig. 3a as a function of pump wavelength and sample tilt angle. The q-BIC $TE_{10}$ exhibits a characteristic "V"-shaped dispersion with enhanced emission at resonance, while the LM $TE_{20}$ exhibits an inverted "Λ"-shaped dispersion and lower nonlinear enhancement, due to its lower $Q$-factor. These observations are consistent with the linear transmission results in Figs. 2a and 2b.

Under TM-polarized excitation, the SHG dispersion features an enhancement with the characteristic "Λ" shape associated with q-BIC $TM_{10}$ along with an inverted "V" shape corresponding to the LM $TM_{20}$ (Fig. 3b). The SHG enhancement by the q-BIC $TM_{10}$ is nearly an order of magnitude lower than that of the $TE_{10}$ mode because the pump polarization is orthogonal to the LN optical axis (see also spectra at fixed tilt angle in panels c and d). This predominantly addresses the weaker $d_{31}$ and $d_{22}$ nonlinear coefficients, which are about one order of magnitude smaller than $d_{33}$ (~4.3 pm/V, ~2.1 pm/V, respectively).

Higher SHG efficiency is expected when optimal coupling condition is met – *i.e.* when the q-BIC $Q$-factor matches the $Q_{laser}$. Indeed, for lower angles of incidence the pump energy cannot be efficiently coupled to the mode due to its large $Q$-factor (> $10^4$). Conversely, for larger angles, although the pump energy coupling is maximized, the reduced $Q$-factor of the mode typically is expected to limit the field enhancement in the LNTF layer. By comparing the experimentally retrieved Q-factors of the q-BIC in Fig. 2j and 2l with the $Q_{laser}$ (see grey shaded region in Fig. 2j), we find that for TE polarization in particular such condition is achieved experimentally at $\theta \sim \pm 0.2°$. The SHG power maps in Figures 3a and 3b reveal an onset of the SHG at $\theta \sim \pm 0.2°$ with a maximum around $\theta \sim \pm 1°$, which is in very good agreement with the optimal coupling prediction. The SHG intensity behavior as a function of the tilt angle for both TE and TM configurations are corroborated by the linear spectra, featuring maximum transmission dips also at $\theta \sim \pm 1°$ as detailed in Figure S5 of the Supplementary Information (SI).

Interestingly, the simulated linear transmission spectra at the SHG wavelength (see Figure

S6 of the SI) reveal the presence of the $TM_{31}$ guided mode resonance intersecting the q-BIC $TM_{10}$ around $\theta = \pm 1$, that also emerges by plotting the SHG maps in logarithmic scale (see Figs. 3f). This additional resonant channel modifies the light–matter interaction at the SHG wavelength, hence producing a modulation of SHG intensity visible around $\theta = \pm 1$ for TM polarization configuration. It is also worth noting that the SHG intensity of the q-BIC $TE_{10}$ mode is almost unchanged for angles up to $\pm 4°$ (see Figure 3c), which allows for robust and precise tuning of the SHG wavelength between 860 nm and 920 nm (*i.e.* 40 nm bandwidth). Finally, the removal of the mirror symmetry induced by the slanting angle of the $TiO_2$ grating allows to sizably couple the picosecond pump source to the q-BIC $TE_{10}$ also at $k_x = 0$, yielding a moderate SHG peak (see Figure S7 of the SI).

We also determined the maximum SHG efficiency normalized to the pump beam intensity $\eta_{I,\text{norm}} = \frac{P_{2\omega}^{\text{avg}}}{P_{\omega}^{\text{avg}}}/I_{\text{peak}}$, where $P_{2\omega}^{\text{avg}}$ and $P_{\omega}^{\text{avg}}$ are the SHG and pump average powers, respectively, and $I_{\text{peak}}$ is the peak intensity of the pump beam (see details in the Supplementary Information S7). This figure of merit rules out the pump spectral characteristics, incident power, and beam size, quantifying the intrinsic second-harmonic conversion capability of the metasurface. The estimated $\eta_{I,\text{norm}} \sim 1.5 \times 10^{-3} \frac{cm^2}{GW} = 0.15\% \frac{cm^2}{GW}$ is among the largest reported in literature for LN nanophotonic platforms[29,30].

Beyond intensity-dependent measurements, we investigated the polarization state of the emitted SHG (see the Methods Section). Figures. 4a and 4b display TE- and TM-polarized SHG maps under TE excitation, plotted on a logarithmic scale to make weak features visible. Under TE excitation, both the q-BIC $TE_{10}$ and the LM $TE_{20}$ exhibit TE-polarized SHG, that is two orders of magnitude stronger than the TM component. This stems from the enhanced fundamental fields being predominantly oriented along the nanograting direction, thereby efficiently probing the $d_{33}$ nonlinear coefficient of LN. In contrast, TM excitation leads to a more complex fundamental field distribution with significant components both along the TM direction and normal to the metasurface. In combination with the smaller non-diagonal elements ($d_{31}$ and $d_{22}$) of the LN nonlinear tensor, this field distribution results in a dominant TM-polarized SHG contribution accompanied by a weaker, yet non-negligible, TE-polarized component, as shown in Figs. 4d and 4e.

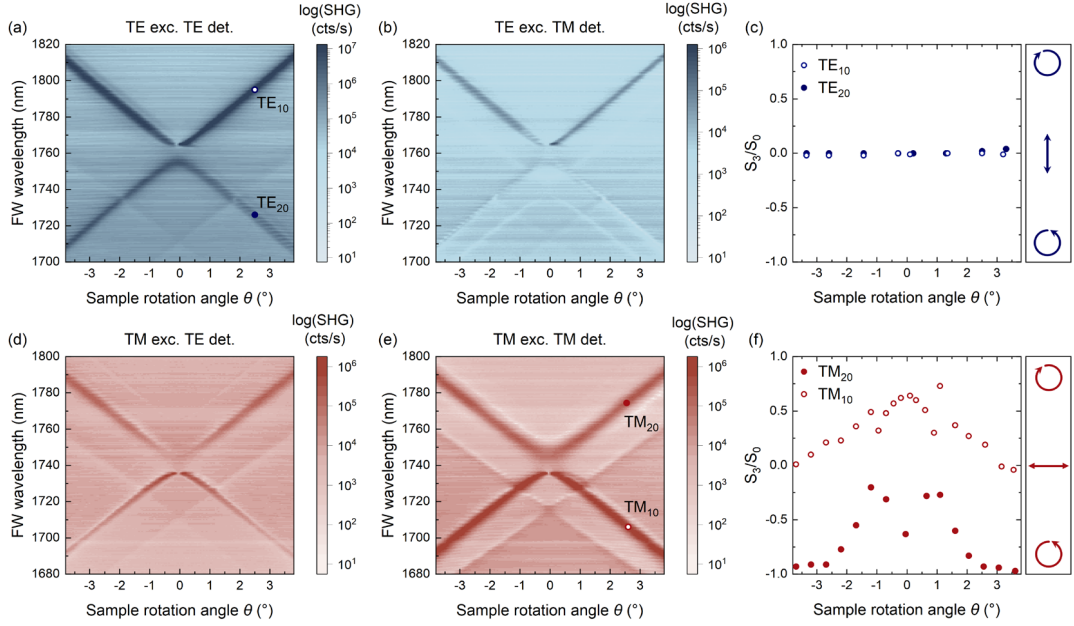

**Figure 4.** Polarization-dependent SHG response of the nanostructured sample. (a,b) Logarithmic SHG spectra under TE-polarized excitation, with TE and TM polarization-sensitive detection, respectively, as a function of the sample rotation angle $\theta$. Note the different color scale range of panel (a). (c) $\theta$-dependence of the normalized Stokes parameter $S_3/S_0$ of the modes $TE_{10}$ and $TE_{20}$. (d,e) Logarithmic SHG spectra under TM-polarized excitation, with detection in TE and TM polarizations, respectively. (f) $\theta$-dependence of the normalized Stokes parameter $S_3/S_0$ of the modes $TM_{20}$ and $TM_{10}$.

To precisely determine the SHG polarization state, we performed full polarimetric measurements and retrieved all the Stokes parameters, following the approach described in Ref. [31]. Under TE excitation, the vanishing normalized Stokes parameter $S_3/S_0$ (Fig. 4c) representing the degree of circular polarization (DOCP) confirms that the SHG from both the q-BIC $TE_{10}$ and the LM $TE_{20}$ is linearly polarized along the $z$ direction at all pump incidence angles. In contrast, the $S_3/S_0$ parameter characterizing the SHG DOCP under TM excitation (Fig. 4f) reveals a pronounced polarization modulation with sample rotation. For the q-BIC $TM_{10}$, the SHG evolves from right-handed elliptical polarization at $\theta \approx 0°$ (DOCP = 63%) to nearly $x$-oriented linear polarization at $\theta \approx \pm 4°$, whereas for the LM $TM_{20}$ mode the polarization changes from nearly $x$-oriented linear polarization at $\theta \approx \pm 1°$ to almost fully circular (DOCP = 97%) at $\theta \approx \pm 4°$. This behavior provides a new avenue for tunable SHG not only in wavelength

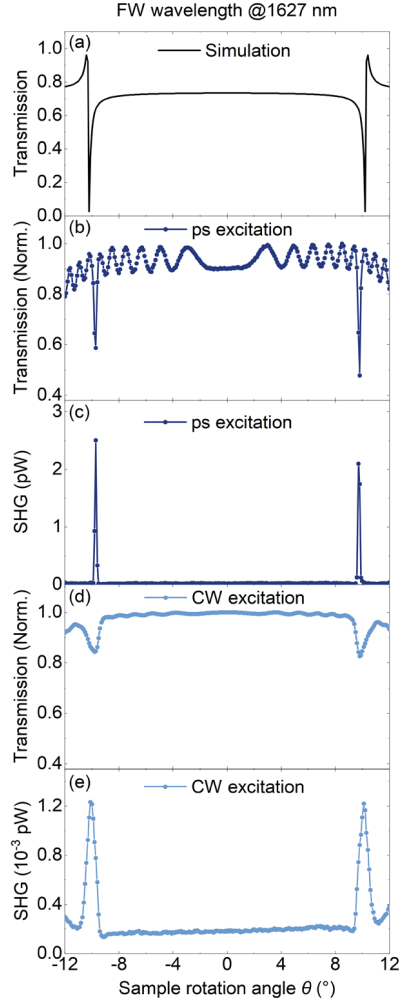

**Figure 5.** Large-angle linear and nonlinear optical responses under TE-polarized excitation at a fixed wavelength of 1627 nm. (a) Simulated transmission spectra as a function of sample tilt angle $\theta$ from −12° to +12°. (b,c) Experimental transmission and SHG measured with a picosecond pump laser and 10 mW average power. (d,e) Experimental transmission and SHG measured with a a continuous-wave laser emitting 10 mW power.

and intensity but also in polarization.

The extremely low peak intensity employed (3.64 kW/cm$^2$) indicates that sizeable SHG is expected also employing continuous wave (CW) laser with moderate power. We verified such claim by exciting the metasurface with a CW laser at a wavelength of about 1627 nm with TE polarization and comparing the results with those obtained with the picosecond laser at the same wavelength. Firstly, we performed angle-resolved transmission measurements with both laser sources and compared them with simulations (Figs. 5a). The angular transmission plots in Figures 5b and 5d allow addressing resonant dips at $\theta \approx \pm 10°$, corresponding to the LM TE$_{20}$ mode. We then performed SHG angular plots with both ps (Fig. 5c) and CW (Fig. 5e) pump laser using the same average power of 10 mW. The estimated SHG power obtained using CW excitation ($\sim 1.2\ fW$) is approximately $2 \times 10^3$ times lower than the one seeded by picosecond pulses ($\sim 2.4\ pW$). This is trivially explained by the fact that the ratio between the SHG power seeded by a pulsed laser ($P_{SHG}^{ps}$) and the one seeded by a CW one ($P_{SHG}^{CW}$) is equal to the inverse of the pulsed laser duty cycle ($D$): $\frac{P_{SHG}^{ps}}{P_{SHG}^{CW}} = D^{-1}$. Considering the employed picosecond laser pulse duration ($\tau_{\text{pulse}} \sim 7\ ps$) and repetition rate ($R = 50\ MHz$), we arrive to $D^{-1} = (\tau_{\text{pulse}} \cdot R)^{-1} \sim 2.8 \times 10^3$, which is in very good agreement with the experimentally-retrieved power ratio.

**Conclusion**

In summary, we have demonstrated a large-area (1 mm × 1 mm) nonlinear photonic platform by integrating a 5° tilted TiO$_2$ nanograting, via nanoimprint lithography, onto an *x*-cut thin film of lithium niobate. By exploiting q-BIC resonances supported by the hybrid grating–film structure, we achieved strong SHG under picosecond-pulsed excitation with relatively low peak intensities. Importantly, by finely tuning the angle between the sample normal and the light propagation direction, we were able to dynamically match the q-BIC linewidth to the laser bandwidth, thereby maximizing light–matter interaction. This approach yields a maximum normalized SHG efficiency of $\eta_{I,\text{norm}} = 0.15\%$ cm$^2$/GW, which is among the largest reported thus far for LN nanoscale devices. In particular, the remarkably low maximum peak intensity employed in this study (3.64 kW/cm$^2$) underscores the potential deployment of this platform in integrated devices and operation with low-power (mW) continuous wave sources. More importantly, the SHG wavelength can be continuously tuned from 870 nm to 920 nm through small angular adjustments, while the SHG intensity remains within the same order of magnitude, across the entire tuning range under TE excitation. Under TM excitation, the platform further enables polarization modulation from linear to circular, offering a versatile degree of control over both the spectral and polarization properties of the emitted SHG. By combining high conversion efficiency with dynamic spectral tunability and polarization-sensitive response, our nanoimprinted LN metasurface emerges as a powerful platform for compact frequency conversion, integrated tunable light sources, and reconfigurable nonlinear photonic systems, uniquely enhanced by the large-area scalability enabled by nanoimprint lithography.

**Acknowledgements**

The authors would like to thank Fabrizio Conti for support during the initial setup of the nonlinear experiments. This work is supported by the Italian Ministry of University and Research (MUR) through the PRIN 2022 project "NoLIMITHz" (Id: 2022BC5BW5; CUP D53D23001130006) and through Project NQSTI—ID PE_00000023 funded by the European Union under the NextGenerationEU program - CUP H43C22000870001 Spoke 6. A.Z., M.F., and M.C.

**Materials and methods**

**Sample fabrication**

The TiO$_2$ nanograting is made of dispersing titanium dioxide nanoparticles in a UV-curable resin, also named TiO$_2$ nanoparticle-embedded-resin (TiO$_2$ nano-PER). The nanogratings were fabricated using an electric-field-assisted nanoimprint lithography (EF-NIL) technique, as described in detail in Ref.[32]. A flexible template replicated from a silicon master mold was conformally brought into contact with the substrate through a line-contact rolling process under an applied electric field. The electric field provided the driving force for resist filling and ensured uniform replication. After UV curing, the flexible template was peeled off to complete the pattern transfer. This approach enables high-fidelity, large-area nanostructure fabrication while maintaining excellent structural uniformity.

**Numerical calculations**

The optical response of the LN metasurface was numerically simulated using the commercial software COMSOL Multiphysics, which is based on the finite-element method. An *x*-cut LN wafer (with the optical axis parallel to the *z*-axis, as shown in Fig. 1b) was employed as the material under study. The optical modes of the device were analyzed using the implemented 2D finite-element eigenfrequency solver. The linear transmittance was obtained by exciting the metasurface with a plane wave from the nanograting side and collecting the transmitted light on the substrate side. The nonlinear optical process was simulated using a two-step cascaded computational approach. In the first step, the linear optical response of the metasurface was calculated at the fundamental wavelength (FW) to obtain the electric field distribution inside the device. The nonlinear polarization distribution inside the LN film oscillating at the second-harmonic frequency was evaluated using the reduced nonlinear tensor, with nonzero elements $d_{22}$ = 2.1 pm/V, $d_{31}$ = −4.3 pm/V and $d_{33}$ = −27 pm/V and served as the source for the second computational step at the SHG frequency.

**Experimental setup**

Linear characterization:

A schematic of the setup is presented in Figure S1 of the SI. Linear transmittance measurements were performed using a supercontinuum laser source (NKT Photonics, SuperK FIANIUM) covering a spectral range from 400 to 2400 nm. The laser output was collimated by a telescope to minimize the angular spread. The sample was illuminated from the nanograting side, and the transmitted light was collected from the substrate side. Angular dependence was assessed by mounting the sample on a motorized rotation stage (Thorlabs, PRM1/MZ8) enabling the measurement of resonance linewidths as a function of the tilt angle. The transmitted signal was analyzed by a spectrometer (Andor, Kymera328i, 400 l/mm grating, 1600 nm blaze) equipped with an InGaAs array (Andor, iDus) yielding a spectral resolution of ~0.5 nm.

Nonlinear characterization:

A sketch of the nonlinear setup is presented in Figure S2 of the SI. We employed a tunable picosecond laser (Stuttgart Instruments, Piano) with repetition rate = 50 MHz, pulse duration ≈ 7 ps, and a constant linewidth of 4 $cm^{-1}$ (about 0.5 nm in the wavelength range of interest) to pump the sample from the nanograting side, while SHG signals were collected from the substrate side. The pump beam was first expanded and collimated with a telescope (Thorlabs, AC254-075-C and AC254-200-C) and cropped by an iris to define a ~1 mm diameter spot, matching the sample area. The CW laser diode source was a fiber-coupled laser diode (Thorlabs LPSC-1625-FC) emitting at 1627 nm. The generated SHG signal was filtered sequentially with an IR-Cut filter (Rapid Spectral Solutions 1540SP) to remove the transmitted pump light and a long-pass filter (Thorlabs, FEL800) to eliminate higher-order spectral components, ensuring a clean SHG output. The SHG signal was then focused by a converging lens (Thorlabs AC254-150-B) and detected using a highly sensitive Si-based single-photon avalanche diode (Micro Photon Devices, PDM Series). For the polarization study, we placed a half-waveplate (Thorlabs, AHWP10M-980) and a linear polarizer (Thorlabs, LPNIR050-MP) for the TE and TM polarized SHG maps and a quarter-waveplate (Thorlabs, AQWP10M-580) and a linear polarizer (Thorlabs, LPNIR050-MP) for evaluating the Stokes parameters.

**References**

[1] L. Bonacina, P.-F. Brevet, M. Finazzi, M. Celebrano, *J. Appl. Phys.* **2020**, *127*, 230901.

[2] N. C. Panoiu, *Fundamentals and Applications of Nonlinear Nanophotonics*, Elsevier, **2023**.

[3] A. Di Francescantonio, A. Zilli, D. Rocco, V. Vinel, L. Coudrat, F. Conti, P. Biagioni, L. Duò, A. Lemaître, C. De Angelis, G. Leo, M. Finazzi, M. Celebrano, *Nat. Nanotechnol.* **2024**, *19*, 298.

[4] A. Sinelnik, S. H. Lam, F. Coviello, S. Klimmer, G. Della Valle, D.-Y. Choi, T. Pertsch, G. Soavi, I. Staude, *Nat. Commun.* **2024**, *15*, 2507.

[5] A. Verneuil, A. Zilli, C. Vézy, J. Béal, M. Finazzi, M. Celebrano, A. Baudrion, *Adv. Opt. Mater.* **2025**, *13*, e01748.

[6] A. Aghigh, S. Bancelin, M. Rivard, M. Pinsard, H. Ibrahim, F. Légaré, *Biophys. Rev.* **2023**, *15*, 43.

[7] A. N. Poddubny, D. N. Neshev, A. A. Sukhorukov, in *Nonlinear Meta-Optics*, **2020**, pp. 147–180.

[8] P. Vabishchevich, Y. Kivshar, *Photon. Res.* **2023**, *11*, B50.

[9] K. Koshelev, S. Kruk, E. Melik-Gaykazyan, J.-H. Choi, A. Bogdanov, H.-G. Park, Y. Kivshar, *Science* **2020**, *367*, 288.

[10] N. C. Panoiu, W. E. I. Sha, D. Y. Lei, G.-C. Li, *J. Opt.* **2018**, *20*, 083001.

[11] M. Celebrano, X. Wu, M. Baselli, S. Großmann, P. Biagioni, A. Locatelli, C. De Angelis, G. Cerullo, R. Osellame, B. Hecht, L. Duò, F. Ciccacci, M. Finazzi, *Nat. Nanotechnol.* **2015**, *10*, 412.

[12] G. Grinblat, *ACS Photonics* **2021**, *8*, 3406.

[13] A. Fedotova, L. Carletti, A. Zilli, F. Setzpfandt, I. Staude, A. Toma, M. Finazzi, C. De Angelis, T. Pertsch, D. N. Neshev, M. Celebrano, *ACS Photonics* **2022**, *9*, 3745.

[14] A. Boes, L. Chang, C. Langrock, M. Yu, M. Zhang, Q. Lin, M. Lončar, M. Fejer, J. Bowers, A. Mitchell, *Science* **2023**, *379*, eabj4396.

[15] A. Fedotova, M. Younesi, J. Sautter, A. Vaskin, F. J. F. Löchner, M. Steinert, R. Geiss, T. Pertsch, I. Staude, F. Setzpfandt, *Nano Lett.* **2020**, *20*, 8608.

[16] L. Carletti, A. Zilli, F. Moia, A. Toma, M. Finazzi, C. De Angelis, D. N. Neshev, M. Celebrano, *ACS Photonics* **2021**, *8*, 731.

[17] K. Koshelev, S. Lepeshov, M. Liu, A. Bogdanov, Y. Kivshar, *Phys. Rev. Lett.* **2018**, *121*, 193903.

[18] R. Kolkowski, T. K. Hakala, A. Shevchenko, M. J. Huttunen, *Appl. Phys. Lett.* **2023**, *122*, 160502.

[19] X. Zhang, L. He, X. Gan, X. Huang, Y. Du, Z. Zhai, Z. Li, Y. Zheng, X. Chen, Y. Cai, X. Ao, *Laser & Photonics Rev.* **2022**, *16*, 2200031.

[20] S. Yuan, Y. Wu, Z. Dang, C. Zeng, X. Qi, G. Guo, X. Ren, J. Xia, *Phys. Rev. Lett.* **2021**, *127*, 153901.

[21] A. Di Francescantonio, A. Sabatti, H. Weigand, E. Bailly-Rioufreyt, M. A. Vincenti, L. Carletti, J. Kellner, A. Zilli, M. Finazzi, M. Celebrano, R. Grange, *Nat. Commun.* **2025**, *16*, 7000.

[22] F. Kaufmann, G. Finco, A. Maeder, R. Grange, *Nanophotonics* **2023**, *12*, 1601.

[23] J. Zhang, J. Ma, M. Parry, M. Cai, R. Camacho-Morales, L. Xu, D. N. Neshev, A. A. Sukhorukov, *Sci. Adv.* **2022**, *8*, eabq4240.

[24] Z. Yu, Y. Tong, H. K. Tsang, X. Sun, *Nat. Commun.* **2020**, *11*, 2602.

[25] F. Ye, Y. Yu, X. Xi, X. Sun, *Laser & Photonics Rev.* **2022**, *16*, 2100429.
[26] F. Gambino, M. Giaquinto, A. Ricciardi, A. Cusano, *Results Opt.* **2022**, *6*, 100210.
[27] J. Kühne, J. Wang, T. Weber, L. Kühner, S. A. Maier, A. Tittl, *Nanophotonics* **2021**, *10*, 4305.
[28] M. M. Choy, R. L. Byer, *Phys. Rev. B* **1976**, *14*, 1693.
[29] L. Qu, Z. Gu, C. Li, Y. Qin, Y. Zhang, D. Zhang, J. Zhao, Q. Liu, C. Jin, L. Wang, W. Wu, W. Cai, H. Liu, M. Ren, J. Xu, *Adv. Funct. Mater.* **2023**, *33*, 2308484.
[30] L. Qu, W. Wu, W. Cai, M. Ren, J. Xu, *Laser & Photonics Reviews* **2025**, *19*, 2401928.
[31] Y. Luan, A. Zilli, A. Di Francescantonio, V. Vinel, P. Biagioni, L. Duò, A. Lemaître, G. Leo, M. Celebrano, M. Finazzi, *Light Sci. Appl.* **2025**, *14*, 318.
[32] Y. Fan, C. Wang, H. Tian, X. Chen, B. Q. Li, Z. Wang, X. Li, X. Chen, J. Shao, *Nano-Micro Lett.* **2026**, *18*, 12.